\documentclass[twocolumn]{aastex6}
\usepackage{epstopdf}
\usepackage{amsmath,array,graphicx}

\usepackage{array,graphicx,tabularx,threeparttable, capt-of}
\usepackage{booktabs}
\usepackage{pifont}
\usepackage{multirow}
\usepackage{natbib}

\fullcollaborationName{The Friends of AASTeX Collaboration}

\begin{document}


\title{Large-scale Control of Kinetic Dissipation in the Solar Wind}


\author{Daniel Vech\altaffilmark{1}, Kristopher G. Klein\altaffilmark{2,1} and Justin C. Kasper\altaffilmark{1},}

\affil{$^{1}$Climate and Space Sciences and Engineering, University of Michigan, Ann Arbor, MI 48109, USA, $^{2}$Lunar and Planetary Laboratory, University of Arizona, Tucson, AZ 85719, USA; dvech@umich.edu}
\begin{abstract}
In this Letter we study the connection between the large-scale dynamics of the turbulence cascade and particle heating on kinetic scales. We find that the inertial range turbulence amplitude ($\delta B_i$; measured in the range of 0.01-0.1 Hz) is a simple and effective proxy to identify the onset of significant ion heating and when it is combined with $\beta_{||p}$, it characterizes the energy partitioning between protons and electrons ($T_p/T_e$), proton temperature anisotropy ($T_{\perp}/T_{||}$) and scalar proton temperature ($T_p$) in a way that is consistent with previous predictions. For a fixed $\delta B_i$, the ratio of linear to nonlinear timescales is strongly correlated with the scalar proton temperature in agreement with Matthaeus et al., though for solar wind intervals with $\beta_{||p}>1$ some discrepancies are found. For a fixed $\beta_{||p}$, an increase of the turbulence amplitude leads to higher $T_p/T_e$ ratios, which is consistent with the models of Chandran et al. and Wu et al. We discuss the implications of these findings for our understanding of plasma turbulence.
\end{abstract}




\section{Introduction}\label{sec:intro}

The solar wind is ubiquitously observed to be in a turbulent state with a power spectrum of fluctuations spanning from magnetohyrodynamic (MHD) to smaller kinetic scales \citep[e.g.][]{coleman1968turbulence, siscoe1968power}. Between approximately $10^{-4}$ and 0.3 Hz in the spacecraft frame the magnetic fluctuations form a power law called the inertial range, which is characterized with a spectral index of -5/3 corresponding to the fluid scaling \citep[e.g.][]{podestaaniso, horbury2012anisotropy}. At approximately $f_{sc}=$ 0.3 Hz the spectrum steepens and energy starts to dissipate leading to particle heating \citep[e.g.][]{leamon1998observational, bruno2005solar, howes2008model}. In the sub-ion range the turbulence has Alfv\'enic nature \citep{chen2013nature} and the spectrum has a clearly non-universal spectral index in the range of -4 to -2 \citep[e.g.][]{leamon1999dissipation, sahraoui2010three}, which has been found to be correlated with the energy cascade rate \citep[e.g.][]{smith2006dependence, matthaeus2008interplanetary}. The partitioning of the dissipated energy between protons and electrons is thought to be affected by several plasma parameters including the nonlinear timescales \citep{matthaeus2016turbulence}, gyroscale turbulence amplitude \citep{chandran2010perpendicular} and the ratio of parallel thermal pressure to magnetic pressure \citep[][]{cerri2017plasma}; ($\beta_{||p}$=$2\mu_0n_p k_BT_{||p}/B_0^2$). 


A crucial factor characterizing the energy cascade rate and the relative heating of protons and electrons is the nonlinear timescale at which the energy is transferred to smaller scales \citep[see review by][]{horbury2012anisotropy}.
\cite{goldreich1995toward} proposed the critical balance theory predicting that the linear timescale corresponding to the propagating Alfv\'enic fluctuations and their nonlinear decay are comparable at each scale: $\tau_{A} (k_{\perp}) \sim \tau_{CB}(k_{\perp})$ where $k_{\perp}$ is the perpendicular (with respect to the magnetic field) wavenumber. The Alfv\'en time and the nonlinear ``critical balance" time are estimated for a given spatial scale perpendicular to the background magnetic field ($\lambda \sim 2\pi/k_{\perp}$) as

\begin{equation}
\tau_A(\lambda) \sim \frac{l_\parallel}{V_A} \sim \left( \frac{ L}{\lambda} \right)^{1/3} \frac{\lambda}{V_A}
\label{eqn:tau_A}
\end{equation}

\begin{equation}
\tau_{CB}(\lambda) \sim \frac{\lambda}{\delta z_{\lambda}}
\label{eqn:tau_CB}
\end{equation}
where $l_{||}$ is the spatial scale along the magnetic field, $V_A$ is the Alfv\'en speed ($B_0/(\rho \mu_0)^{1/2}$), L is the size of the outer scale of the cascade and $\delta z = \delta v + \delta b$, $\delta v$, and $\delta b$ are the Els{\"a}sser, velocity and magnetic fluctuations at scale $\lambda$, respectively. The perpendicular scale of the eddies decreases at a faster rate than the parallel scale, with the scaling $k_\parallel \propto k^{2/3}_{\perp}$. Both observational \citep[e.g.][]{horbury2008anisotropic, podestaaniso, wicks2010power, Chen:2011a} and numerical \citep[e.g.][]{cho2000anisotropy, maron2001simulations, TenBarge:2012a} studies are consistent with critical balance scalings; see \cite{Chen:2016} for a detailed review.


In contrast to critical balance theory, \cite{matthaeus2014nonlinear} argued that the most significant contributions to nonlinear spectral transfer are independent of $\tau_A$. They proposed that at kinetic scales the relevant time scale ratio is between the gyroperiod,$\tau_{ci}\sim \Omega_{ci}^{-1}$, and nonlinear turnover time (e.g. the time it takes until an eddy at scale $l$ passes all of its energy to a smaller scale) at scale $l \sim d_i$,
\begin{equation}
\tau_{nl} (d_i) = \frac{V_A}{Z \Omega_{ci}} \left(\frac{L}{d_i}  \right)^{1/3},
\label{eqn:tau_nl}
\end{equation}
which determines how the dissipated energy is partitioned in proton and electron heating ($Q_p/Q_e \sim 1 / (\tau_{nl} \Omega_{ci})$). 
The ion inertial length is $d_i =c/\omega_{pi}$, $\omega_{pi} = \sqrt[]{n_i q_i^2 / \epsilon_0m_{i}}$ is the ion plasma frequency, $\Omega_{ci} = qB_0/m$ is the proton gyrofrequency. The total energy per unit mass is given as $Z = \sqrt[]{ u^2 + b^2}$ where u and b denote the root-mean-square velocity and magnetic field fluctuations, the latter measured in velocity units ($b=b_{rms}/\sqrt[]{\mu_0 \rho}$).

In addition to the nonlinear timescales, the magnitude of the gyroscale velocity fluctuations also plays an important role in controlling the proton and electron heating. When the electromagnetic field fluctuations at gyroscale surpass a critical amplitude the first adiabatic invariant of particle motion is not conserved allowing perpendicular heating of the particles known as stochastic ion heating \citep[e.g.][]{mcchesney1987observation, johnson2001stochastic}. \cite{chandran2010perpendicular} proposed that stochastic heating depends on the dimensionless parameter $\epsilon=\delta v_{\rho}/ v_{\perp} $ where $\delta v_{\rho}$ is the root-mean-square velocity fluctuations at gyroscale and $v_{\perp}$ is the ion's thermal speed perpendicular to the magnetic field. The perpendicular proton heating rate per unit mass ($Q_{\perp}$) at $k_{\perp} \rho_p \sim 1$ (where $\rho_p=v_{th\perp i}/\Omega_i$ is the proton gyroscale) as a fraction of the turbulent cascade power per unit mass ($\Gamma$), assuming a balanced spectrum of kinetic Alfv\'en waves (KAWs) can be given in the form of
\begin{equation}
\frac{Q_{\perp}}{\Gamma} = 3.0 \exp\left(\frac{-0.34}{\epsilon}\right).
\label{eqn:q_perp}
\end{equation}
Equation $\eqref{eqn:q_perp}$ implies that half of the total cascade power is directed into perpendicular proton heating at $k_{\perp} \rho_p \sim 1$ when $\epsilon = 0.19$. 

Another significant parameter affecting the dissipation process is $\beta_{||p}$, which enhances or completely restricts the operation of certain heating mechanisms. When $\beta_{||p} \ll 1$ electron Landau damping dominates while proton Landau damping is negligible since the thermal ions are too slow to satisfy the Landau resonance condition \citep{quataert1998particle}. On the other hand, when $\beta_{||p} \sim 1$ Landau and transit time damping of kinetic Alfv\'en waves lead to significant parallel proton heating \citep{gary2004kinetic}. Heating due to reconnection may also depend on $\beta_{||p}$: \cite{mistry2017statistical} found that the temperature increase of the exhaust region is a function of the inflow $\beta_{||p}$ and reconnection guide field. The onset of stochastic heating is thought to be independent of $\beta_{||p}$ for $\beta_{||p} \lesssim 1$ \citep{chandran2010perpendicular}. 

Electron and proton heating by solar wind turbulence have been investigated by both observational \citep[e.g.][]{cranmer2009empirical, coburn2012turbulent, he2015evidence,  sorriso2018local} and numerical studies \citep[e.g.][]{breech2009electron, servidio2012local, wan2015intermittent, wan2016intermittency, gary2016whistler}. \cite{wu2013karman} used particle-in-cell simulation in the presence of a strong magnetic field to study how the decaying energy in the turbulent cascade is partitioned between protons and electrons and concluded that as the turbulence energy increases protons are heated more. The crossover value ($T_e = T_p$) occurred when the initial turbulence amplitude ($\delta b/B_0$) reached 2/5. They suggested that the correlation between the proton heating and turbulence amplitude is primarily due to the increased involvement of coherent structures in the kinetic processes \citep[e.g.][]{parashar2009kinetic, markovskii2010effect, greco2012inhomogeneous}.

\cite{cerri2017plasma} compared a two-dimensional (2-D) hybrid Vlasov-Maxwell simulation of externally driven turbulence and a hybrid 2-D particle-in-cell simulation of freely decaying turbulence. Despite the fundamental differences between the two simulations, the kinetic scale turbulence was remarkably similar: the root-mean-square amplitudes of the density, parallel and perpendicular magnetic field fluctuations showed less than a factor of two difference and depended only on $\beta$. \cite{cerri2017plasma} concluded that regardless how the large-scale fluctuations are injected, the system continuously ``reprocesses" the turbulent fluctuations as they are cascading towards smaller scales and the response of the system is primarily driven by $\beta$.

In this Letter, we continue this general line of inquiry and study how the dissipated energy is partitioned between protons and electrons in the solar wind as a function of the strength of the cascade. To quantify the strength of the cascade we use a directly measurable proxy, the average inertial range amplitude ($\delta B_i$) of the turbulence spectrum of magnetic fluctuations. We find that the ($\beta_{||p}$, $\delta B_i$) space organizes the solar wind plasma measurements in a way that is consistent with current theories about solar wind heating in particular with \cite{chandran2010perpendicular}, \cite{wu2013karman}, \cite{matthaeus2016turbulence}, and characterizes the proton-electron temperature ratio ($T_p/T_e$), proton temperature anisotropy ($T_{\perp}/T_{||}$) and scalar proton temperature ($T_p$). Finally, we aim to identify the timescale ratio that has the best correlation with $T_p$. For this purpose we test $\tau_{ci} / \tau_{nl}(d_i)$, $\tau_A(\rho_p) / \tau_{CB}(\rho_p)$ and a ``hybrid" timescale ratio defined as $\tau_A(\rho_p) / \tau_{CB}(\rho_p) \times  \exp(-0.34/\epsilon)$ incorporating the effect of stochastic ion heating.

\section{Method} \label{sec:results}
We selected Wind magnetic field \citep{lepping1995wind} (92 ms cadence), ion (SWE FC, 92 second cadence) and electron (45 second cadence) data \citep{lin1995three, ogilvie1995swe} from January 2004 to December 2016 and split the time series into 10-minute intervals. For each of the $\sim 5.8 \cdot 10^5$ intervals $T_{||}$, $T_{\perp}$ (with orientations defined based on the average magnetic field direction during each 92 second interval), $\beta_{||p}$ and $T_e$ were averaged. The power spectral density (PSD) of the magnetic field components were calculated separately via Fourier transform and the component PSDs were summed up to obtain the total PSD \citep{koval2013magnetic}. The spectral index in the inertial range was calculated by fitting the PSD between $0.01 - 0.1$ Hz; $\delta B_i$ corresponds to the average (in log space) power level measured in this frequency range. The average and standard deviation of the measured spectral indices are -1.68 $\pm$ 0.26, respectively in excellent agreement with previous studies \citep[e.g. ][]{leamon1999dissipation, smith2006dependence, alexandrova2009universality}.


To estimate $\tau_A$ (Equation $\ref{eqn:tau_A}$) and $\tau_{nl}$ (Equation $\ref{eqn:tau_nl}$) we assume that the spectral break between the outer and inertial ranges of the turbulence cascade is at a constant frequency of $10^{-4}$ Hz \citep[e.g.][]{podestaaniso, wicks2011anisotropy} and calculate the size of the outer scale L as $V_{sw}/(2 \pi 10^{-4})$ where $V_{sw}$ is the solar wind speed. \cite{matthaeus2014nonlinear} suggested that under typical solar wind conditions $Z/V_A$ is expected to be in the range of 0.5-1. We calculated Z based on the root-mean-square velocity and magnetic field fluctuations during each 10 min interval and found that the median $Z/V_A$ ratio is 0.42.

Measuring the gyroscale velocity fluctuations with current instruments is only possible under exceptional solar wind conditions. To be able to conduct a statistical study we use the approach of \cite{bourouaine2013observational} to estimate $\delta v_{\rho}$ in Equation $\eqref{eqn:tau_CB}$ based on the spectrum of magnetic field fluctuations as $\delta v_{\rho} = \sigma V_A \delta B / B_0$ where $\sigma = 1.19$ is a dimensionless constant arising from the kinetic Alfv\'en dispersion relation and $\delta B$ is the gyroscale turbulence amplitude. For details of the technique and its application for a statistical study see \cite{bourouaine2013observational} and \cite{vech2017nature}. For the calculation of $\delta b$ in Equation $\eqref{eqn:tau_CB}$ we used the gyroscale turbulence amplitude expressed in Alfv\'en units: $\delta b = \delta B / (\mu_0 \rho)^{1/2}$.

\section{Results} \label{sec:res}

\begin{figure}
\figurenum{1}
    \centering\includegraphics[width=1\linewidth]{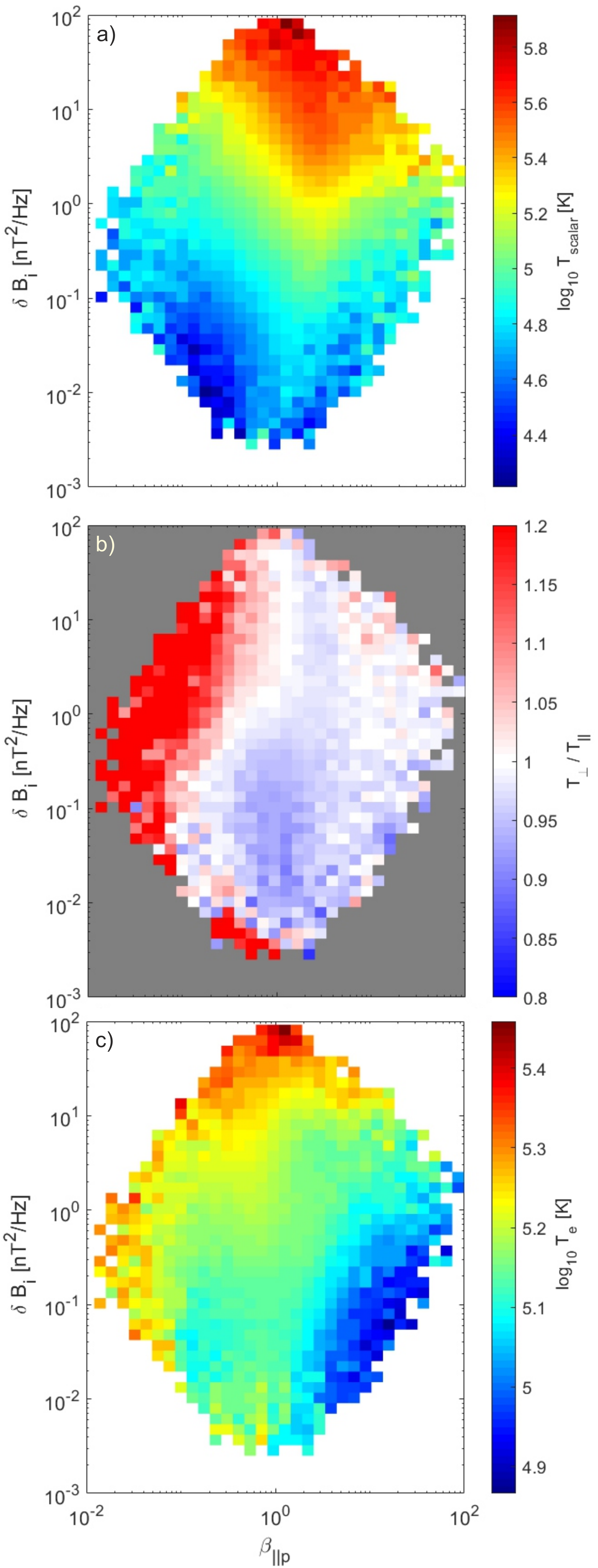}
    \caption{Median values of the scalar proton temperature (a), proton temperature anisotropy (b) and electron temperature (c) in the ($\beta_{||p}$, $\delta B_i$) space.}
    \label{fig:1}
    \end{figure}	
      
    \begin{figure}
\figurenum{2}
    \centering\includegraphics[width=1\linewidth]{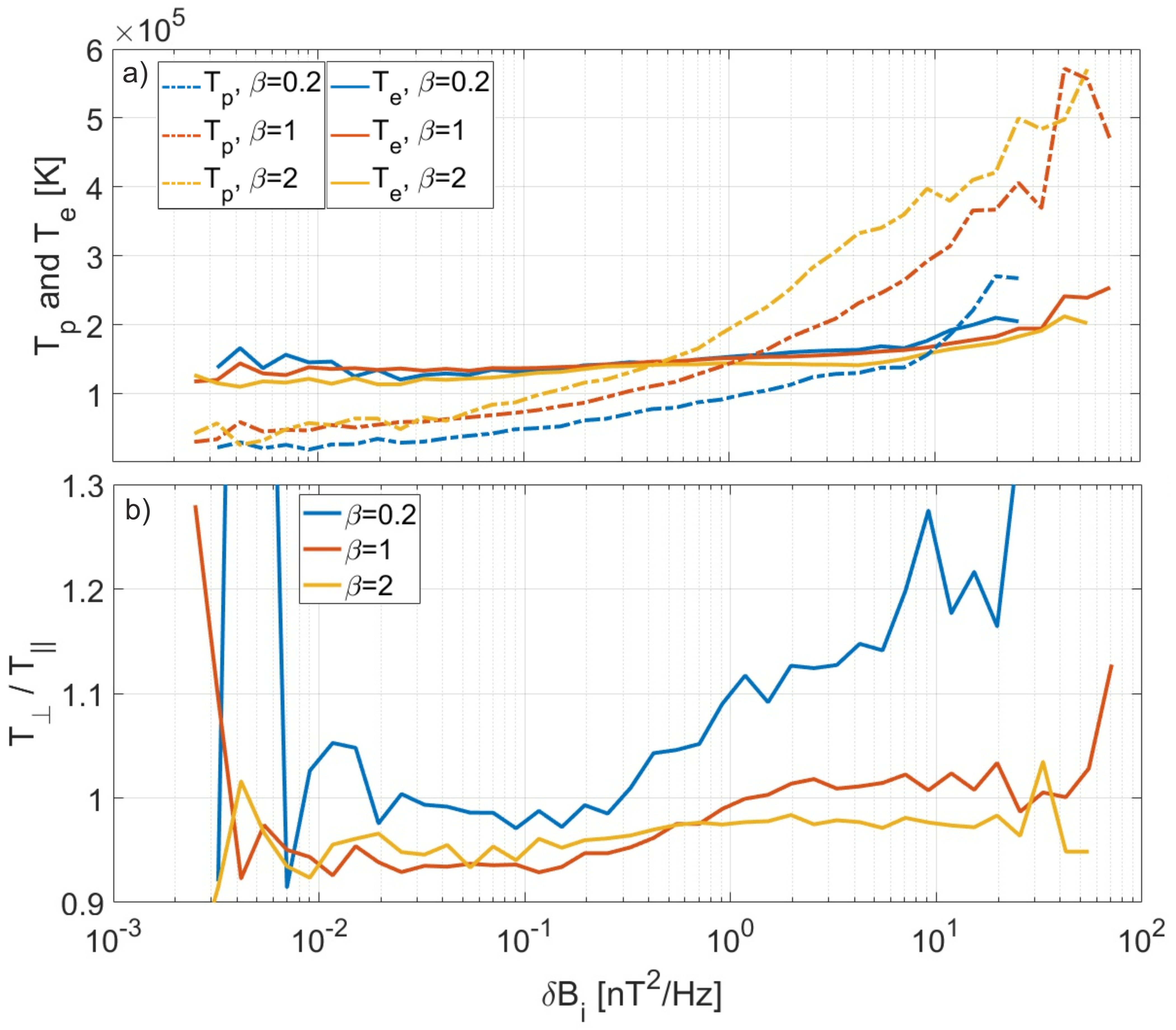}
    \caption{a) Cross sections of Figures \ref{fig:1}a (dashed lines) and c (solid lines) along $\beta_{||p}=$0.2, 1 and 2, respectively. b) Cross section of Figure \ref{fig:1}b along $\beta_{||p}=$0.2, 1 and 2, respectively.}
    \label{fig:2}
    \end{figure}

    \begin{figure}
\figurenum{3}
    \centering\includegraphics[width=1\linewidth]{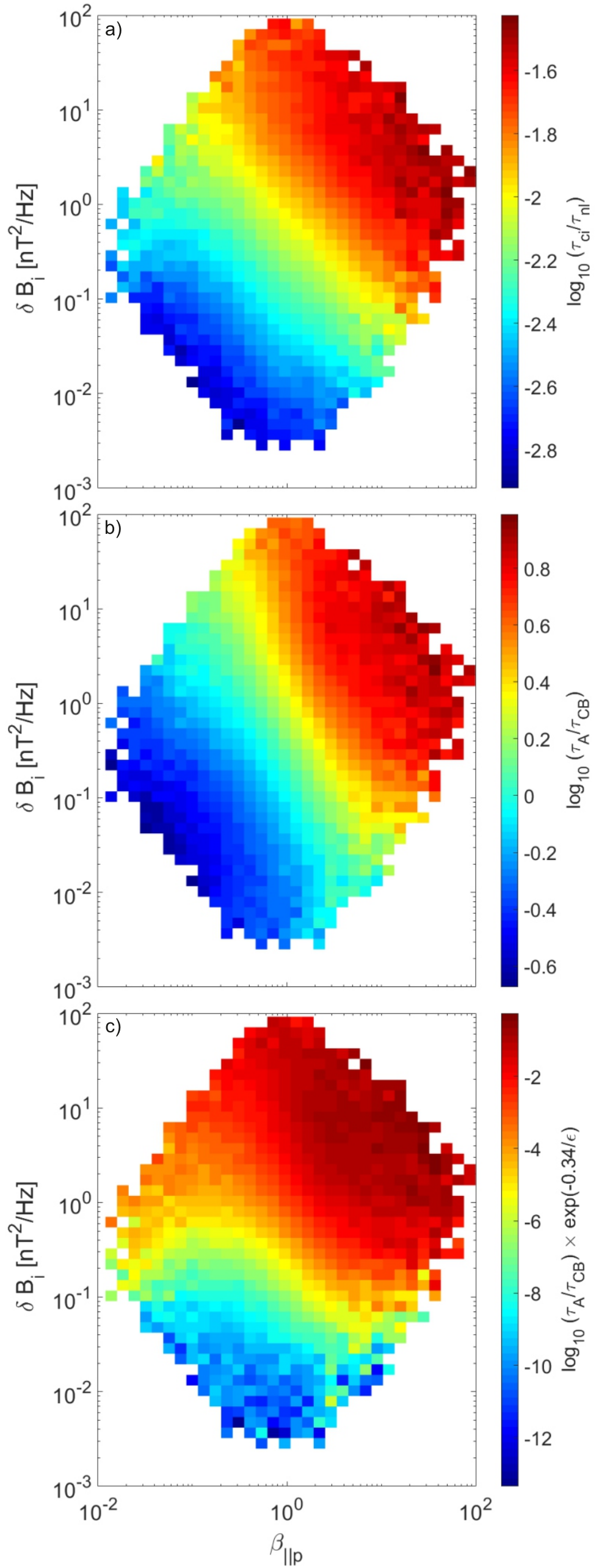}
    \caption{Median values of $\tau_{ci}/\tau_{nl}(d_i)$ (a), $\tau_A /\tau_{CB}(\rho_p)$ (b), $\tau_A /\tau_{CB}(\rho_p) \times exp(-0.34/\epsilon)$ (c) in the ($\beta_{||p}$, $\delta B_i$) space.}
    \label{fig:3}
    \end{figure}

    \begin{figure}
\figurenum{4}
    \centering\includegraphics[width=1\linewidth]{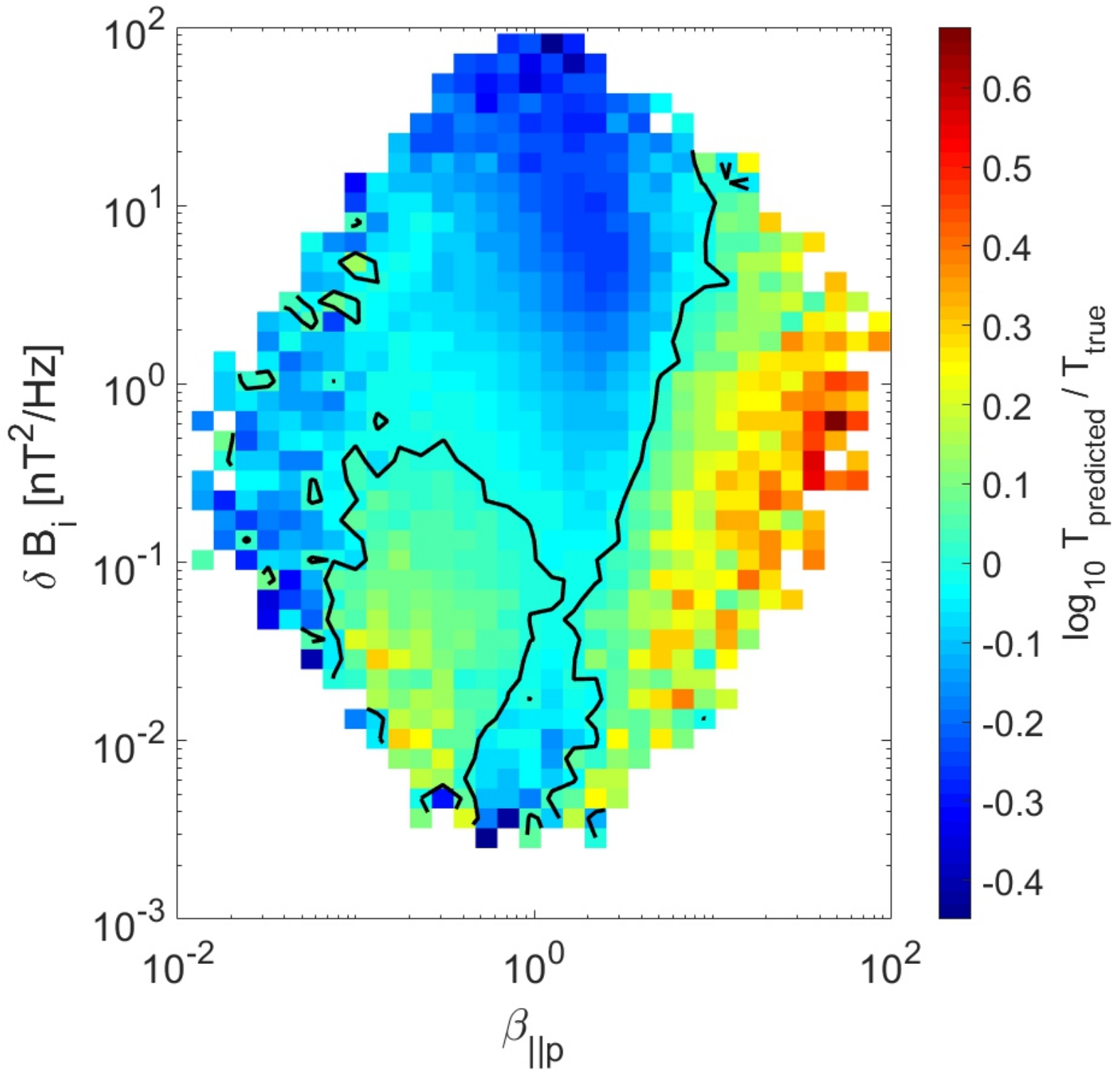}
    \caption{Median values of the $T_{predicted}/T_{true}$ ratios in the ($\beta_{||p}$, $\delta B_i$) space testing the $\beta_{||p} \delta B_i \sim T_p$ scaling. The contour indicates the $T_{true}=T_{predicted}$ boundary.}
    \label{fig:4}
    \end{figure}

The distributions of $T_p$, $T_{\perp}/T_{||}$ and $T_e$ were studied in 2-D histograms with 50x50 logarithmically spaced bins in the ($\beta_{||p}, \delta B_i$) space. The median of each bin was selected and sparse bins with fewer than 10 data points were discarded. In our data set the medians of $\beta_{||p}$ and $\delta B_i$ are $0.99$ and $0.72$ nT$^2$/Hz, respectively.
 
The scalar proton temperature in Figure~\ref{fig:1}a increases as a function of $\delta B_i$ and when $\delta B_i$ is larger than 0.2 nT$^2$/Hz the peak temperature is around $4 \cdot 10^5$ K while for $\delta B_i < 0.1$ nT$^2$/Hz the temperature is around $3 \cdot 10^4$ K. The $\delta B_i$ dependence of the scalar proton temperature is shown in Figure~\ref{fig:2}a for three values of $\beta_{\parallel p}$ as dashed lines: in all cases the temperature increases nearly exponentially as a function of log$_{10} \delta B_i$.

In Figure~\ref{fig:1}b, the proton temperature anisotropy is significantly different for the $\beta_{||p}<1$ and $\beta_{||p}>1$ regions: for small $\beta_{||p}$ the anisotropy increases as a function of $\delta B_i$ while for large $\beta_{||p}$ no obvious systematic trend can be seen. In Figure ~\ref{fig:2}b, the temperature anisotropy for $\beta_{||p}=0.2$ is nearly constant when $\delta B_i<0.2 $ nT$^2$/Hz while for $\delta B_i > 0.2$ nT$^2$/Hz there is a clear indication of perpendicular proton heating ($T_{\perp}/T_{||} \sim 1.15$). In the case of $\beta_{||p}$=1, for low $\delta B_i$ values minor parallel heating is observed ($T_{\perp}/T_{||} \sim 0.95$) and the temperature anisotropy reaches unity in the high $\delta B_i$ limit.

In Figure~\ref{fig:1}c the $T_e$ distribution shows positive correlation with $\delta B_i$ and the $T_e$ values change with approximately a factor of three across the whole range of $\delta B_i$. The solid lines in Figure ~\ref{fig:2}a show the $T_e$ cross-sections for the same values of $(\beta_{\parallel p}$. When $\delta B_i <$0.1 nT$^2$/ Hz, $T_p/T_e < 1$ while for the largest $\delta B_i$ values the protons have a factor of 2.5 higher temperature than electrons. It is important to note that the proton-electron temperature equilibrium shows significant dependence on $\beta_{||p}$: as $\beta_{||p}$ changes from 0.2 to 2 the $\delta B_i$ value corresponding to $T_p=T_e$ significantly decreases meaning that in a plasma with high $\beta_{||p}$ even relatively small magnetic fluctuations are sufficient to produce equal proton and electron temperatures.

In Figure~\ref{fig:3} we calculate the value of three timescale ratios, $\tau_{ci} / \tau_{nl}(d_i)$, $\tau_A(\rho_p) / \tau_{CB}(\rho_p)$ and a ``hybrid" timescale ratio defined as $\tau_A(\rho_p) / \tau_{CB}(\rho_p) \times  \exp(-0.34/\epsilon)$  in the $(\beta_{||p}, \delta B_i)$ space. As expected, the distribution of all three ratios show similarities to the proton temperature shown in Figure~\ref{fig:1}a. The correlation between Figure~\ref{fig:3} and~\ref{fig:1}a is weakest in the region where $\delta B_i \in$ [0.01;0.3] nT$^2$/Hz and $\beta_{||p}>1$. For a fixed $\beta_{||p}$ value there is positive correlation between proton temperature and the time scale ratios, which is in qualitative agreement with the prediction of \cite{matthaeus2016turbulence}. To estimate the uncertainties in Figure~\ref{fig:3} we computed the ratio of the standard deviation and mean in each bin. The average uncertainties are 29\%, 25\%, 34\% for Figure~\ref{fig:3}a,b,c, respectively. We note that the errors are the lowest (below 10\%) for $\beta_{p||}>1$ and $\delta B_i > 0.1$ nT$^2$/Hz.

To quantify the correlation between the three distributions in Figure~\ref{fig:3} with Figure~\ref{fig:1}a we use the Spearman's rank correlation ($R_S$), which measures how well the relationship between the time scale ratios and $T_p$ can be described with a monotonic function. The correlations between the binned time scale ratios and $T_p$ are $R_s=0.83, 0.82, 0.88$ for $\tau_{ci}/\tau_{nl}$, $\tau_A /\tau_{CB}$ and $\tau_A /\tau_{CB} \exp(-0.34/\epsilon)$, respectively. We are therefore unable to distinguish between the predictive power of these timescale ratios in determining $T_p$.


As $T_p \sim \beta_{\parallel p} B^2 \sim \beta_{\parallel p}\delta B_i$, the proton temperature distribution may simply be a linear function of the abscissa and ordinate variables of Figure~\ref{fig:1}. To test this dependence we binned $0.03 \times \beta_{\parallel p} \delta B_i$ as a function of the ($\beta_{||p}, \delta B_i$); the factor of 0.03 Hz corresponds to the center (in log space) of the frequency range where $\delta B_i$ is measured. The values of $0.03 \times \beta_{\parallel p} \delta B_i$ data were multiplied with a constant factor of $5.9 \times 10^5$ so it had the same mean as the mean of the observed proton temperature. Finally, a least-square fit ($y=0.201x+4.258$) was made between the logarithm of $0.03 \times \beta_{\parallel p} \delta B_i$ (x) and logarithm of the actual proton temperature (y) data. If the linear $\beta_{\parallel p}$ and $\delta B_i$ dependencies are the only significant factors in the behavior of $T_p$ then we expect that the predicted proton temperature ($T_{predicted}$) based on the power law fit to agree well with the the observed proton temperature ($T_{true}$). Figure~\ref{fig:4} shows the ratio $T_{predicted}/T_{true}$ in the ($\beta_{||p}, \delta B_i$) space. Three major features can be observed: for $\beta_{||p}<1$ and $\delta B_i<0.2 $ nT$^2$/Hz the observed proton temperature is lower than the predicted values with a factor of 1.5 while for $\delta B_i>0.2 $ nT$^2$/Hz the observed temperature is higher by a factor of two. The discrepancy is the most significant for high $\beta_{||p}$ where the observed temperature is a factor of three lower than the predicted one. This is also the region with the lowest correlation between the timescale ratios and $T_p$. Therefore we conclude that the naive $T_p \sim \beta_{||p} \delta B_i$ scaling is not sufficient to explain the variability of the $T_p$ distribution in Figure~\ref{fig:1}a.

\section{Conclusion} \label{sec:conclusion}

In this Letter we have studied the connection between the inertial range of the turbulent cascade and the small scale dissipation in the solar wind as function of the inertial range turbulence amplitude $\delta B_i$ and $\beta_{||p}$. Our approach links directly the characteristics of the turbulence spectrum of magnetic fluctuations to heating mechanisms on kinetic scales therefore it could be potentially a simple and effective tool to diagnose heating in the solar wind and in plasma systems more generally.

\cite{vech2017nature} identified the onset of stochastic heating when $\epsilon = \delta v_{\rho}/v_{\perp}$ reached 0.025 and 76\% of the studied intervals had an $\epsilon$ value larger than this. Here we used the exact same time interval allowing a direct comparison between $\epsilon$ and $\delta B_i$: when $\delta B_i$ is in the range of $0.1-0.3$ nT$^2$/Hz (e.g. approximately where the sudden perpendicular temperature enhancement is observed in Figure~\ref{fig:2}b) the median $\epsilon$ is 0.029 and 74\% of the intervals had an $\delta B_i$ value larger than $0.3$ nT$^2$/Hz. Due to this excellent agreement between critical values of $\epsilon$ and $\delta B_i$ we interpret the sudden enhancement of $T_{\perp}/T_{\parallel }$ as the onset of stochastic ion heating. The evolution of the temperature parameters in Figures~\ref{fig:1}-\ref{fig:2} across a critical threshold of turbulence amplitude is in qualitative agreement with the stochastic ion heating model of \cite{chandran2010perpendicular}.

We note that the model of \cite{chandran2010perpendicular} was parameterized for gyroscale velocity fluctuations to identify the critical turbulence amplitude when the gyromotion of protons is disrupted. Our findings suggest that stochastic heating is controlled by large-scale dynamics of the turbulent cascade and reaching the critical turbulence amplitude at gyroscale is a direct consequence of the increased energy cascade rate from larger scales.

For a fixed $\beta_{||p}$, the $T_p/T_e$ ratio increases as a function of $\delta B_i$. When the turbulence amplitude is small ($\delta B_i < 0.3$ nT$^2$/Hz) electrons are hotter, while for larger turbulence amplitudes $T_p/T_e > 1$. As $\beta_{||p}$ increases $T_p=T_e$ occurs at smaller $\delta B_i$ meaning that in a high $\beta_{||p}$ plasma even relatively small turbulence amplitudes can lead to equal proton and electron temperatures. These findings may be especially relevant for astrophysical plasmas where $\beta_p \gg 1$. Our results are in qualitative agreement with the predictions of \cite{wu2013karman}, however we note that the increased proton temperatures as a function of $\delta B_i$ may be partially caused by stochastic ion heating, the effects of coherent structures in the proton heating, or both mechanisms.

The timescale ratios had similar distributions in the ($\beta_{||p}, \delta B_i$) space and they all had strong correlation with the proton temperature data ($0.88 > R_S >0.82$), thus in our data they are indistinguishable. For a fixed $\delta B_i$ value, the $T_p/T_e$ ratio increases as a function of all the timescale ratios, which is consistent with the prediction of \cite{matthaeus2016turbulence}. The weakest correlation between the timescale ratios and $T_p$ was observed for high $\beta_{||p}$.

Finally, \cite{cerri2017plasma} suggested that the response of a plasma system is primarily driven by the amount of available energy at kinetic scales and $\beta_{||p}$. Our findings are in agreement with this concept and $\beta_{||p}$ may have the most significant influence on the proton-electron temperature ratio by restricting and enhancing the operation of certain heating mechanisms.

\acknowledgments

K.G. Klein was supported by NASA grant NNX16AM23G. J.C. Kasper was supported by NASA Grant NNX14AR78G. Data were sourced from CDAWeb (http://cdaweb.gsfc.nasa.gov/).


\begin{thebibliography}{}
\expandafter\ifx\csname natexlab\endcsname\relax\def\natexlab#1{#1}\fi

\bibitem[{Alexandrova {et~al.}(2009)Alexandrova, Saur, Lacombe, Mangeney,
  Mitchell, Schwartz, \& Robert}]{alexandrova2009universality}
Alexandrova, O., Saur, J., Lacombe, C., {et~al.} 2009, Physical review letters,
  103, 165003

\bibitem[{Bourouaine \& Chandran(2013)}]{bourouaine2013observational}
Bourouaine, S., \& Chandran, B.~D. 2013, The Astrophysical Journal, 774, 96

\bibitem[{Breech {et~al.}(2009)Breech, Matthaeus, Cranmer, Kasper, \&
  Oughton}]{breech2009electron}
Breech, B., Matthaeus, W.~H., Cranmer, S., Kasper, J., \& Oughton, S. 2009,
  Journal of Geophysical Research: Space Physics, 114

\bibitem[{Bruno \& Carbone(2005)}]{bruno2005solar}
Bruno, R., \& Carbone, V. 2005, Living Reviews in Solar Physics, 2, 4

\bibitem[{Cerri {et~al.}(2017)Cerri, Franci, Califano, Landi, \&
  Hellinger}]{cerri2017plasma}
Cerri, S., Franci, L., Califano, F., Landi, S., \& Hellinger, P. 2017, Journal
  of Plasma Physics, 83

\bibitem[{Chandran {et~al.}(2010)Chandran, Li, Rogers, Quataert, \&
  Germaschewski}]{chandran2010perpendicular}
Chandran, B.~D., Li, B., Rogers, B.~N., Quataert, E., \& Germaschewski, K.
  2010, The Astrophysical Journal, 720, 503

\bibitem[{Chen {et~al.}(2013)Chen, Boldyrev, Xia, \& Perez}]{chen2013nature}
Chen, C., Boldyrev, S., Xia, Q., \& Perez, J. 2013, Physical review letters,
  110, 225002

\bibitem[{{Chen}(2016)}]{Chen:2016}
{Chen}, C.~H.~K. 2016, Journal of Plasma Physics, 82, 535820602

\bibitem[{{Chen} {et~al.}(2011){Chen}, {Mallet}, {Yousef}, {Schekochihin}, \&
  {Horbury}}]{Chen:2011a}
{Chen}, C.~H.~K., {Mallet}, A., {Yousef}, T.~A., {Schekochihin}, A.~A., \&
  {Horbury}, T.~S. 2011, 415, 3219

\bibitem[{Cho \& Vishniac(2000)}]{cho2000anisotropy}
Cho, J., \& Vishniac, E.~T. 2000, The Astrophysical Journal, 539, 273

\bibitem[{Coburn {et~al.}(2012)Coburn, Smith, Vasquez, Stawarz, \&
  Forman}]{coburn2012turbulent}
Coburn, J.~T., Smith, C.~W., Vasquez, B.~J., Stawarz, J.~E., \& Forman, M.~A.
  2012, The Astrophysical Journal, 754, 93

\bibitem[{Coleman~Jr(1968)}]{coleman1968turbulence}
Coleman~Jr, P.~J. 1968, The Astrophysical Journal, 153, 371

\bibitem[{Cranmer {et~al.}(2009)Cranmer, Matthaeus, Breech, \&
  Kasper}]{cranmer2009empirical}
Cranmer, S.~R., Matthaeus, W.~H., Breech, B.~A., \& Kasper, J.~C. 2009, The
  Astrophysical Journal, 702, 1604

\bibitem[{Gary {et~al.}(2016)Gary, Hughes, \& Wang}]{gary2016whistler}
Gary, S.~P., Hughes, R.~S., \& Wang, J. 2016, The Astrophysical Journal, 816,
  102

\bibitem[{Gary \& Nishimura(2004)}]{gary2004kinetic}
Gary, S.~P., \& Nishimura, K. 2004, Journal of Geophysical Research: Space
  Physics, 109

\bibitem[{Goldreich \& Sridhar(1995)}]{goldreich1995toward}
Goldreich, P., \& Sridhar, S. 1995, The Astrophysical Journal, 438, 763

\bibitem[{Greco {et~al.}(2012)Greco, Valentini, Servidio, \&
  Matthaeus}]{greco2012inhomogeneous}
Greco, A., Valentini, F., Servidio, S., \& Matthaeus, W. 2012, Physical Review
  E, 86, 066405

\bibitem[{He {et~al.}(2015)He, Wang, Tu, Marsch, \& Zong}]{he2015evidence}
He, J., Wang, L., Tu, C., Marsch, E., \& Zong, Q. 2015, The Astrophysical
  Journal Letters, 800, L31

\bibitem[{Horbury {et~al.}(2012)Horbury, Wicks, \&
  Chen}]{horbury2012anisotropy}
Horbury, T., Wicks, R., \& Chen, C. 2012, Space Science Reviews, 172, 325

\bibitem[{Horbury {et~al.}(2008)Horbury, Forman, \&
  Oughton}]{horbury2008anisotropic}
Horbury, T.~S., Forman, M., \& Oughton, S. 2008, Physical Review Letters, 101,
  175005

\bibitem[{Howes {et~al.}(2008)Howes, Cowley, Dorland, Hammett, Quataert, \&
  Schekochihin}]{howes2008model}
Howes, G.~G., Cowley, S.~C., Dorland, W., {et~al.} 2008, Journal of Geophysical
  Research: Space Physics, 113

\bibitem[{Johnson \& Cheng(2001)}]{johnson2001stochastic}
Johnson, J.~R., \& Cheng, C. 2001, Geophysical research letters, 28, 4421

\bibitem[{Koval \& Szabo(2013)}]{koval2013magnetic}
Koval, A., \& Szabo, A. 2013in , AIP, 211--214

\bibitem[{Leamon {et~al.}(1998)Leamon, Smith, Ness, Matthaeus, \&
  Wong}]{leamon1998observational}
Leamon, R.~J., Smith, C.~W., Ness, N.~F., Matthaeus, W.~H., \& Wong, H.~K.
  1998, Journal of Geophysical Research: Space Physics, 103, 4775

\bibitem[{Leamon {et~al.}(1999)Leamon, Smith, Ness, \&
  Wong}]{leamon1999dissipation}
Leamon, R.~J., Smith, C.~W., Ness, N.~F., \& Wong, H.~K. 1999, Journal of
  Geophysical Research: Space Physics, 104, 22331

\bibitem[{Lepping {et~al.}(1995)Lepping, Ac{\~u}na, Burlaga, Farrell, Slavin,
  Schatten, Mariani, Ness, Neubauer, Whang, {et~al.}}]{lepping1995wind}
Lepping, R., Ac{\~u}na, M., Burlaga, L., {et~al.} 1995, Space Science Reviews,
  71, 207

\bibitem[{Lin {et~al.}(1995)Lin, Anderson, Ashford, Carlson, Curtis, Ergun,
  Larson, McFadden, McCarthy, Parks, {et~al.}}]{lin1995three}
Lin, R., Anderson, K., Ashford, S., {et~al.} 1995, Space Science Reviews, 71,
  125

\bibitem[{Markovskii \& Vasquez(2010)}]{markovskii2010effect}
Markovskii, S., \& Vasquez, B.~J. 2010, Physics of Plasmas, 17, 112902

\bibitem[{Maron \& Goldreich(2001)}]{maron2001simulations}
Maron, J., \& Goldreich, P. 2001, The Astrophysical Journal, 554, 1175

\bibitem[{Matthaeus {et~al.}(2008)Matthaeus, Weygand, Chuychai, Dasso, Smith,
  \& Kivelson}]{matthaeus2008interplanetary}
Matthaeus, W., Weygand, J., Chuychai, P., {et~al.} 2008, The Astrophysical
  Journal Letters, 678, L141

\bibitem[{Matthaeus {et~al.}(2016)Matthaeus, Parashar, Wan, \&
  Wu}]{matthaeus2016turbulence}
Matthaeus, W.~H., Parashar, T.~N., Wan, M., \& Wu, P. 2016, The Astrophysical
  Journal Letters, 827, L7

\bibitem[{Matthaeus {et~al.}(2014)Matthaeus, Oughton, Osman, Servidio, Wan,
  Gary, Shay, Valentini, Roytershteyn, Karimabadi,
  {et~al.}}]{matthaeus2014nonlinear}
Matthaeus, W.~H., Oughton, S., Osman, K.~T., {et~al.} 2014, The Astrophysical
  Journal, 790, 155

\bibitem[{McChesney {et~al.}(1987)McChesney, Stern, \&
  Bellan}]{mcchesney1987observation}
McChesney, J., Stern, R., \& Bellan, P. 1987, Physical Review Letters, 59, 1436

\bibitem[{Mistry {et~al.}(2017)Mistry, Eastwood, Phan, \&
  Hietala}]{mistry2017statistical}
Mistry, R., Eastwood, J., Phan, T., \& Hietala, H. 2017, Journal of Geophysical
  Research: Space Physics

\bibitem[{Ogilvie {et~al.}(1995)Ogilvie, Chornay, Fritzenreiter, Hunsaker,
  Keller, Lobell, Miller, Scudder, Sittler, Torbert, {et~al.}}]{ogilvie1995swe}
Ogilvie, K., Chornay, D., Fritzenreiter, R., {et~al.} 1995, Space Science
  Reviews, 71, 55

\bibitem[{Parashar {et~al.}(2009)Parashar, Shay, Cassak, \&
  Matthaeus}]{parashar2009kinetic}
Parashar, T., Shay, M., Cassak, P., \& Matthaeus, W. 2009, Physics of Plasmas,
  16, 032310

\bibitem[{{Podesta}(2009)}]{podestaaniso}
{Podesta}, J.~J. 2009, Astrophys. J., 698, 986

\bibitem[{Quataert(1998)}]{quataert1998particle}
Quataert, E. 1998, The Astrophysical Journal, 500, 978

\bibitem[{Sahraoui {et~al.}(2010)Sahraoui, Goldstein, Belmont, Canu, \&
  Rezeau}]{sahraoui2010three}
Sahraoui, F., Goldstein, M.~L., Belmont, G., Canu, P., \& Rezeau, L. 2010,
  Physical review letters, 105, 131101

\bibitem[{Servidio {et~al.}(2012)Servidio, Valentini, Califano, \&
  Veltri}]{servidio2012local}
Servidio, S., Valentini, F., Califano, F., \& Veltri, P. 2012, Physical review
  letters, 108, 045001

\bibitem[{Siscoe {et~al.}(1968)Siscoe, Davis, Coleman, Smith, \&
  Jones}]{siscoe1968power}
Siscoe, G., Davis, L., Coleman, P., Smith, E., \& Jones, D. 1968, Journal of
  Geophysical Research, 73, 61

\bibitem[{Smith {et~al.}(2006)Smith, Hamilton, Vasquez, \&
  Leamon}]{smith2006dependence}
Smith, C.~W., Hamilton, K., Vasquez, B.~J., \& Leamon, R.~J. 2006, The
  Astrophysical Journal Letters, 645, L85

\bibitem[{Sorriso-Valvo {et~al.}(2018)Sorriso-Valvo, Carbone, Perri, Greco,
  Marino, \& Bruno}]{sorriso2018local}
Sorriso-Valvo, L., Carbone, F., Perri, S., {et~al.} 2018, Solar Physics, 293,
  10

\bibitem[{{TenBarge} \& {Howes}(2012)}]{TenBarge:2012a}
{TenBarge}, J.~M., \& {Howes}, G.~G. 2012, 19, 055901

\bibitem[{Vech {et~al.}(2017)Vech, Klein, \& Kasper}]{vech2017nature}
Vech, D., Klein, K.~G., \& Kasper, J.~C. 2017, The Astrophysical Journal
  Letters, 850, L11

\bibitem[{Wan {et~al.}(2015)Wan, Matthaeus, Roytershteyn, Karimabadi, Parashar,
  Wu, \& Shay}]{wan2015intermittent}
Wan, M., Matthaeus, W., Roytershteyn, V., {et~al.} 2015, Physical review
  letters, 114, 175002

\bibitem[{Wan {et~al.}(2016)Wan, Matthaeus, Roytershteyn, Parashar, Wu, \&
  Karimabadi}]{wan2016intermittency}
---. 2016, Physics of Plasmas, 23, 042307

\bibitem[{Wicks {et~al.}(2010)Wicks, Horbury, Chen, \&
  Schekochihin}]{wicks2010power}
Wicks, R., Horbury, T., Chen, C., \& Schekochihin, A. 2010, Monthly Notices of
  the Royal Astronomical Society: Letters, 407, L31

\bibitem[{Wicks {et~al.}(2011)Wicks, Horbury, Chen, \&
  Schekochihin}]{wicks2011anisotropy}
---. 2011, Physical review letters, 106, 045001

\bibitem[{Wu {et~al.}(2013)Wu, Wan, Matthaeus, Shay, \& Swisdak}]{wu2013karman}
Wu, P., Wan, M., Matthaeus, W., Shay, M., \& Swisdak, M. 2013, Physical review
  letters, 111, 121105

\end{thebibliography}
\end{document}